\documentclass{INTERSPEECH2023}
\usepackage{multirow}

\interspeechcameraready

\title{Contextualized End-to-End Speech Recognition with Contextual Phrase Prediction Network}
\name{Kaixun Huang, Ao Zhang, Zhanheng Yang, Pengcheng Guo, Bingshen Mu, Tianyi Xu, Lei Xie$^{*}$\thanks{* Corresponding author.}}
\address{
  Audio, Speech and Language Processing Group (ASLP@NPU), \\ Northwestern Polytechnical University, Xi'an, China}
\email{huangkaixun@mail.nwpu.edu.cn, lxie@nwpu.edu.cn}

\begin{document}

\maketitle

\begin{abstract}
Contextual information plays a crucial role in speech recognition technologies and incorporating it into the end-to-end speech recognition models has drawn immense interest recently. However, previous deep bias methods lacked explicit supervision for bias tasks. In this study, we introduce a contextual phrase prediction network for an attention-based deep bias method. This network predicts context phrases in utterances using contextual embeddings and calculates bias loss to assist in the training of the contextualized model. Our method achieved a significant word error rate (WER) reduction across various end-to-end speech recognition models. Experiments on the LibriSpeech corpus show that our proposed model obtains a 12.1\% relative WER improvement over the baseline model, and the WER of the context phrases decreases relatively by 40.5\%. Moreover, by applying a context phrase filtering strategy, we also effectively eliminate the WER degradation when using a larger biasing list.

\end{abstract}

\vspace{-0.1cm}
\section{Introduction}

\begin{figure*}[t]
\centering
\resizebox{0.8\textwidth}{!}{\includegraphics{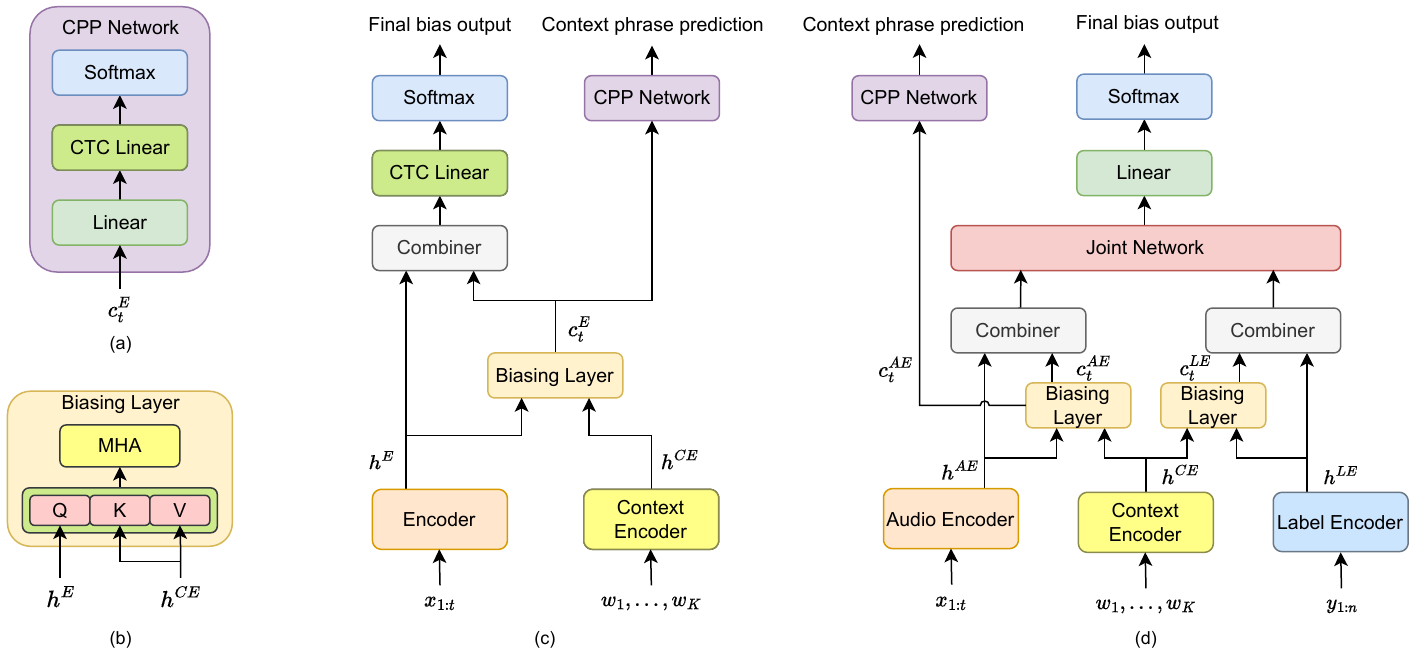}}
\vspace{-0.2cm}
\caption{(a) The proposed contextual phrase prediction network, where the CTC linear shares parameters with contextualized models. (b) Biasing layer via multi-head attention (MHA) (c) Contextualized CTC model (d) Contextualized Transducer model}
\vspace{-0.5cm}
\end{figure*}

In recent years, end-to-end automatic speech recognition (ASR) systems have made remarkable progress, driven by the rapid development of deep neural networks, and have become the mainstream of speech recognition research. Various end-to-end ASR methods, such as connectionist temporal classification (CTC)~\cite{2006ctc, 2014ctc}, recurrent neural network transducer (RNN-T)~\cite{2019rnnt, 2021rnnt}, and attention-based encoder-decoder (AED)~\cite{2015aed, 2014aed, 2016aed}, have been widely used in scenarios such as voice assistants and online conferences.
However, since deep learning techniques are highly data-dependent, the recognition accuracy of end-to-end ASR systems tends to be significantly reduced for rare phrases in the data, such as entity names, technical terms, etc.
In practical applications, these rare phrases can often be obtained in advance as contextual information. For example, for voice assistants, \textit{contact names in the address book} and \textit{the names of installed applications} are likely to appear during the recognition process. 
Therefore, informing the model of contextual information is crucial to accurately predict the corresponding phrases. As a result, research on infusing contextual knowledge into end-to-end ASR models for biased decoding has become increasingly important.

In previous studies, a typical bias method is LM fusion~\cite{2018sf, 2019sf, 2019sf2, 2020sf, 2021sf}, with shallow fusion being the most commonly used approach.
In~\cite{2019sf}, an n-gram of contextual phrases is compiled into a weighted finite-state transducer (WFST) and the state of the decoded path in the WFST is maintained during decoding.
When a specific word is decoded, the score is added to the path, thereby biasing the recognition process toward the contextual phrase.
While LM fusion can intuitively increase the posterior of tokens related to contextual phrases, it has limited performance improvement on the bias list and requires testing to determine the optimal fusion weight.

Another approach to address this problem is the deep bias method based on neural networks~\cite{2019nnbias, 2021nnbias, 2021nnbias2, 2021nnbias3, 2022nnbias}, which introduces a separate biasing component in the end-to-end model for modeling contextual phrases.
Compared with LM fusion, the deep bias method is more flexible and allows quick adaptation to different contextual phrase lists for various scenarios.
Several previous works have modified basic end-to-end speech recognition models to obtain contextualized models, such as CLAS~\cite{2018nnbias} and CATT~\cite{2021nnbias4}. However, these methods lack explicit supervision for bias tasks and are often limited to specific model structures.

To solve this problem, we propose a novel approach called contextual phrase prediction network~(CPP Network) for deep biasing, which combines a bidirectional long-short term memory network (BLSTM) based contextual encoder with an attention mechanism to extract embeddings of context phrases.
Unlike CLAS~\cite{2018nnbias} and CATT~\cite{2021nnbias4}, our model predicts the biased phrases that appear in the utterance using the contextual phrase prediction network and calculates a CTC loss between the predicted phases with labels that only contain context phrases as the bias loss.
This loss function provides explicit supervision for the bias task, allowing the contextualized model to learn to focus on context phrases appearing in the audio while minimizing the impact on the original recognition results when there are no context phrases. The contextual phrase prediction network can be easily integrated into mainstream end-to-end speech recognition models, such as CTC, AED, and Transducer, to achieve contextualized recognition.
Our proposed techniques achieve an average relative WER improvement of 12.1\% on LibriSpeech, with a relative improvement of 40.5\% in the WER of phrases included in the biasing list.

In addition, using a biasing list containing thousands of contextual phrases as input for a contextualized ASR model poses challenges in retrieving meaningful contextual encodings from the biasing layer.
Moreover, the biasing layer requires computations on thousands of contextual phrase encodings, leading to a noticeable increase in inference time. Inspired by~\cite{filter}, we employ a two-stage contextual phrase filtering method.
Firstly, the model performs inference using an empty biasing list to calculate posterior score confidence and sequence order confidence for each context phrase in the actual biasing list. Then, we use the filtered biasing list to generate the final bias output by combining it with the audio embeddings retained from the first stage through a small amount of computation.
This approach reduces the computational burden and inference time, while ensuring that only relevant contextual phrases are used for biasing. Compared to~\cite{filter}, we extended this contextual phrase filtering method to be applicable to multiple structures, rather than being limited to Transducer models with cascaded encoders.

\vspace{-0.15cm}
\section{Contextual Bias with CPP Network}
In this section, we introduce the modification on contextual bias module with our proposed CPP Network. Different modifications made on different ASR framework including CTC, AED and Transducer will be introduced respectively.

\vspace{-0.15cm}
\subsection{Contextualized  CTC/AED}
\vspace{-0.05cm}

\begin{table*}[h]
\centering
\setlength{\belowcaptionskip}{0.05cm} 
\setlength{\abovecaptionskip}{0.15cm} 
\caption{LibriSpeech results with different biasing list size N. Reported metrics are in the following format: WER (U-WER/B-WER).}
\label{librispeech}
\resizebox{0.95\textwidth}{!}{
\begin{tabular}{l|cc|cc|cc|cc}
\hline
\multirow{2}{*}{\qquad \ \ \ Model} & \multicolumn{2}{c|}{N=0}                                                                                               & \multicolumn{2}{c|}{N=100}                                                                                                               & \multicolumn{2}{c|}{N=500}                                                                                              & \multicolumn{2}{c}{N=1000}                                                                                             \\ \cline{2-9} 
                       & test-clean                                                & test-other                                                 & test-clean                                                         & test-other                                                          & test-clean                                                & test-other                                                  & test-clean                                                & test-other                                                  \\ \hline\hline
CTC Baseline           & \begin{tabular}[c]{@{}c@{}}5.59\\ (3.5 / 23.4)\end{tabular} & \begin{tabular}[c]{@{}c@{}}12.17\\ (8.1 / 48.3)\end{tabular} & \begin{tabular}[c]{@{}c@{}}5.59\\ (3.5 / 23.4)\end{tabular}          & \begin{tabular}[c]{@{}c@{}}12.17\\ (8.1 / 48.3)\end{tabular}          & \begin{tabular}[c]{@{}c@{}}5.59\\ (3.5 / 23.4)\end{tabular} & \begin{tabular}[c]{@{}c@{}}12.17\\ (8.1 / 48.3)\end{tabular}  & \begin{tabular}[c]{@{}c@{}}5.59\\ (3.5 / 23.4)\end{tabular} & \begin{tabular}[c]{@{}c@{}}12.17\\ (8.1 / 48.3)\end{tabular}  \\ \hline
CTC + bias     & \begin{tabular}[c]{@{}c@{}}5.67\\ (3.6 / 23.4)\end{tabular} & \begin{tabular}[c]{@{}c@{}}12.14\\ (8.2 / 47.3)\end{tabular} & \textbf{\begin{tabular}[c]{@{}c@{}}5.13\\ (4.2 / 12.7)\end{tabular}} & \textbf{\begin{tabular}[c]{@{}c@{}}11.76\\ (9.9 / 29.0)\end{tabular}} & \begin{tabular}[c]{@{}c@{}}6.11\\ (5.2 / 13.7)\end{tabular} & \begin{tabular}[c]{@{}c@{}}13.77\\ (11.8 / 31.5)\end{tabular} & \begin{tabular}[c]{@{}c@{}}6.89\\ (6.0 / 14.7)\end{tabular} & \begin{tabular}[c]{@{}c@{}}14.85\\ (12.9 / 32.6)\end{tabular} \\ \hline\hline
AED Baseline           & \begin{tabular}[c]{@{}c@{}}4.22\\ (2.6 / 18.1)\end{tabular} & \begin{tabular}[c]{@{}c@{}}8.88\\ (5.6 / 37.6)\end{tabular}  & \begin{tabular}[c]{@{}c@{}}4.22\\ (2.6 / 18.1)\end{tabular}          & \begin{tabular}[c]{@{}c@{}}8.88\\ (5.6 / 37.6)\end{tabular}           & \begin{tabular}[c]{@{}c@{}}4.22\\ (2.6 / 18.1)\end{tabular} & \begin{tabular}[c]{@{}c@{}}8.88\\ (5.6 / 37.6)\end{tabular}   & \begin{tabular}[c]{@{}c@{}}4.22\\ (2.6 / 18.1)\end{tabular} & \begin{tabular}[c]{@{}c@{}}8.88\\ (5.6 / 37.6)\end{tabular}   \\ \hline
AED + bias     & \begin{tabular}[c]{@{}c@{}}4.29\\ (2.6 / 18.3)\end{tabular} & \begin{tabular}[c]{@{}c@{}}9.16\\ (5.9 / 37.5)\end{tabular}  & \textbf{\begin{tabular}[c]{@{}c@{}}3.40\\ (2.6 / 10.4)\end{tabular}}  & \textbf{\begin{tabular}[c]{@{}c@{}}7.77\\ (6.0 / 23.0)\end{tabular}}  & \begin{tabular}[c]{@{}c@{}}3.68\\ (2.8 / 10.9)\end{tabular} & \begin{tabular}[c]{@{}c@{}}8.31\\ (6.5 / 24.3)\end{tabular}   & \begin{tabular}[c]{@{}c@{}}3.81\\ (2.9 / 11.4)\end{tabular} & \begin{tabular}[c]{@{}c@{}}8.75\\ (6.9 / 25.3)\end{tabular}   \\ \hline\hline
Transducer Baseline           & \begin{tabular}[c]{@{}c@{}}4.30\\ (2.8 / 17.2)\end{tabular} & \begin{tabular}[c]{@{}c@{}}8.90\\ (5.9 / 35.5)\end{tabular}  & \begin{tabular}[c]{@{}c@{}}4.30\\ (2.8 / 17.2)\end{tabular}          & \begin{tabular}[c]{@{}c@{}}8.90\\ (5.9 / 35.5)\end{tabular}           & \begin{tabular}[c]{@{}c@{}}4.30\\ (2.8 / 17.2)\end{tabular} & \begin{tabular}[c]{@{}c@{}}8.90\\ (5.9 / 35.5)\end{tabular}   & \begin{tabular}[c]{@{}c@{}}4.30\\ (2.8 / 17.2)\end{tabular} & \begin{tabular}[c]{@{}c@{}}8.90\\ (5.9 / 35.5)\end{tabular}   \\ \hline
Transducer + bias     & \begin{tabular}[c]{@{}c@{}}4.38\\ (2.7 / 18.3)\end{tabular} & \begin{tabular}[c]{@{}c@{}}9.12\\ (5.9 / 37.6)\end{tabular}  & \textbf{\begin{tabular}[c]{@{}c@{}}3.66\\ (2.8 / 11.2)\end{tabular}} & \textbf{\begin{tabular}[c]{@{}c@{}}7.63\\ (6.0 / 22.1)\end{tabular}}  & \begin{tabular}[c]{@{}c@{}}3.78\\ (2.9 / 11.5)\end{tabular} & \begin{tabular}[c]{@{}c@{}}7.99\\ (6.2 / 23.4)\end{tabular}   & \begin{tabular}[c]{@{}c@{}}3.88\\ (2.9 / 11.9)\end{tabular} & \begin{tabular}[c]{@{}c@{}}8.28\\ (6.4 / 24.5)\end{tabular}   \\ \hline
\end{tabular}
}
\vspace{-0.3cm} 
\end{table*}

For both contextualized CTC and AED frameworks, we made modifications only to the encoder. It should be noted that we also deploy CTC as auxiliary training criterion for AED encoder. Thus these modifications for both frameworks were similar. Generally, we add a context encoder, a multi-head attention-based biasing layer, and a contextual phrase prediction network~(CPP Network) for the framework to integrate contextual information, as shown in Figure 1(c).

The context encoder is used to encode the custom contextual phrase list into embeddings with the same length. Each phrase in the list is tokenized with the same tokenizer as the base CTC model, and the token sequence is used as input for the BLSTM. The hidden state and cell state of the last time step from both the forward and backward LSTM are concatenated and then passed through a linear layer to obtain the embedding $h^{CE}_i$ for the contextual phrase. 
In addition, we added a phrase to the contextual phrase list that contains only a single blank token, referred to as \textit{no-bias}, to enable the model to focus on it when there is no contextual phrase in the audio and ignore other contextual phrases.

The biasing layer (Figure 1(b)) is a multi-head attention~(MHA) layer designed to help the model learn the relationship between contextual phrases and speech. While computing the MHA score, the audio embedding $h_t^E$ serves as the query, and the contextual phrase embeddings $h_i^{CE}$ serve as the keys and values to obtain the context representations $c_{t}^E$ as:
\vspace{-0.2cm}
\begin{equation}
\boldsymbol{\alpha}_{t}=\operatorname{Softmax}\left(\left(\mathbf{W}_{q} \mathbf{h}_{t}^{E}\right)\left(\mathbf{W}_{k} \mathbf{h}^{CE}\right)^T / \operatorname{sqrt}\left(d\right)\right)~,
\end{equation}
\begin{equation}
\boldsymbol{c}_{t}^E=\sum_{i=0}^{K} \boldsymbol{\alpha}_{i t} \mathbf{W}_{v} \mathbf{h}_{i}^{CE}~,
\end{equation}
where the scaling factor $\operatorname{sqrt}\left(d\right)$ is for numerical stability, and $K$ represents the number of contextual phrases. The context embeddings and audio embeddings are then combined to obtain context-aware audio embeddings $h_t^{CA}$. The combiner consists of a LayerNorm layer, a concatenation operation, and a feed-forward projection layer. This process can be described as:
\begin{equation}
\mathbf{h}^{concat}_t=[\mathrm{LayerNorm}(\mathbf{h}^{E}_t),\mathrm{LayerNorm}(\boldsymbol{c}^{t})]~,
\end{equation}
\begin{equation}
\mathbf{h}^{CA}_t=\mathrm{FeedForward}(\mathbf{h}^{concat}_t)~.
\end{equation}

The CPP network consists of two feed-forward projection layers and a Softmax function (Figure 1(a)), where the second feed-forward projection layer shares parameters with the linear layer in the CTC model that maps audio embeddings to a posterior probability distribution over vocabularies, denoted as CTC linear in Figure 1.
Receiving the context embedding $c_t^E$ as input, CPP Network predicts the contextual phrases that appear in the dialogue utterance.
We use the posterior generated by CPP Network as input and the contextual phrases in the transcript as reference for CTC loss calculation.
Such loss can supervise the biasing layer to extract relevant contextual phrases information from the audio embeddings $h^E$ when the contextual phrases appear and minimizes the influence of introducing additional information when no contextual phrases appear. 
Note that this part of the model is only used for auxiliary training and does not participate in model inference.

The final joint loss function for the CTC model is:
\vspace{-0.05cm}
\begin{equation}
\setlength\abovedisplayskip{6pt}
\setlength\belowdisplayskip{6pt}
L= L_{ctc}+\lambda_2 L_{CPP}~,
\end{equation}
where $\lambda_2$ is a weight hyperparameter we set as $0.1$ in experiment.

For the AED model, with $\lambda_1$ as the CTC weight, the final joint loss function is:
\vspace{-0.05cm}
\begin{equation}
\setlength\abovedisplayskip{6pt}
\setlength\belowdisplayskip{6pt}
L=\lambda_1 L_{ctc}+(1-\lambda_1) L_{att}+\lambda_1\lambda_2 L_{CPP}~.
\end{equation}

\vspace{-0.3cm}
\subsection{Contextualized Transducer}
\vspace{-0.05cm}

For the Transducer, we augment the architecture with a contextual phrase prediction network to implement the contextualized Transducer, as illustrated in Figure 1(d). Specifically, compared to the standard Transducer, we add a context encoder, two MHA biasing layers, and a contextual phrase prediction network. Similar to CTC/AED framework, both the audio encoder and the label encoder are followed by MHA biasing layers to integrate contextual information from $h_i^{CE}$. Then the context-aware audio embeddings and the context-aware label embeddings are fed into the joint network.



For CPP Network, We use the context embeddings $c_t^{AE}$ obtained from the audio encoder as input to predict the contextual phrases contained in the speech and compute the bias loss.

Finally, we combine the bias loss with the standard Transducer loss as the optimization objective:
\vspace{-0.05cm}
\begin{equation}
\setlength\abovedisplayskip{6pt}
\setlength\belowdisplayskip{6pt}
L=\lambda_1 L_{ctc}+(1-\lambda_1) L_{transducer}+\lambda_1\lambda_2 L_{CPP}~,
\end{equation}
where $\lambda_1$ and $\lambda_2$ are the weight hyperparameters for the CTC loss and the CPP loss.

\vspace{-0.1cm}
\section{Contextual Phrase Filtering}

Inspired by~\cite{filter}, we apply a two-stage filtering algorithm, using the posterior of the CTC to filter the contextual phrases. Differently, we simplified the structure of the Cascade Encoders, performing contextual phrase filtering without altering the original model structure.

 Firstly, the calculation is performed using a biasing list that contains only \textit{no-bias} to obtain the CTC posterior without bias. The audio embedding $h^E$ is retained. The decoding results from the CTC during the first pass inference can serve as streaming results.

The contextual phrases are then filtered in two stages using the CTC posterior. The first stage calculates the posterior and the phrase score confidence (PSC), ignoring the order of the contextual phrases. It computes the sum of the maximum posterior values of each token contained in the contextual phrases within a sliding window and normalizes it with the sequence length. This stage eliminates most of the contextual phrases with low correlation with the audio, retaining only the remaining ones for the second stage of filtering. In the second stage, the sequence order confidence (SOC) is calculated by using a dynamic programming algorithm to compute the maximum posterior sum of tokens contained in contextual phrases within the sliding window, considering token order. 


After obtaining the filtered contextual phrases, subsequent inference is performed with the previously retained audio embedding $h^E$, resulting in a more accurate biased recognition result. The complexity of context phrase filtering is low for inference. Compared to the original recognition process, only the biasing layer, combiner, and CTC linear are repeated in the calculation, and the increase in computation is relatively small.

\section{Experiments}

\vspace{-0.1cm}
\subsection{Data}
\vspace{-0.05cm}

A majority of our experiments were conducted on the LibriSpeech corpus~\cite{librispeech}, which comprises 960 hours of English audiobook readings. The model was trained on the complete 960 hours of training data to demonstrate its performance on relatively low-resource tasks. The \textit{dev-clean} and \textit{dev-other} sets were used for validation, while the \textit{test-clean} and \textit{test-other} sets were used for evaluation. SpecAugment~\cite{specaugment} was applied for data augmentation.

During the training of contextualized models, we extract three random phrases of word length between 1 to 3 from each utterance and add them to the biasing list in the batch. Additionally, we incorporate a certain number of distractors into the biasing list to maintain the list size at 60.

During testing, we utilized the biasing lists for the LibriSpeech corpus provided in~\cite{2021nnbias3} as the contextual phrase lists. As outlined in~\cite{2021nnbias3}, the biasing lists were first formed for each utterance by identifying words belonging to the full rare-word list from the reference of each utterance, and adding a certain number of distractor words. 

To validate the model's performance with larger amounts of data, we also evaluate our method on GigaSpeech corpus~\cite{gigaspeech} and tested on \textit{Earnings21} dataset~\cite{earnings}. \textit{Earnings21} set consists of earnings call transcripts from various publicly traded companies and provides a range of proprietary terms, including names of companies, products, and executives.

\vspace{-0.15cm}
\subsection{Experimental Setups}
\vspace{-0.05cm}

The baseline CTC model consists of a 12-layer conformer with 256-dimensional input and 4 self-attention heads and the baseline AED model uses 12 conformer layers for the encoder and 6 transformer layers for the decoder. The baseline Transducer model uses a 12-layer conformer as the audio encoder and 2-layer LSTM as the label encoder, referred to as C-T.


In the contextualized version, the context encoder consists of a 1-layer BLSTM with a dimensionality of 256 and a linear layer, while the biasing layer is an MHA layer with an embedding size of 256 and 4 attention heads. The first linear layer in the context prediction network maintains a 256-dimensional input and output, followed by a tanh activation function, while the second linear layer projects the input onto the vocabulary size and shares parameters with the CTC linear.

In addition to the word error rate (WER), we also evaluated the results using the biased word error rate (B-WER) and unbiased word error rate (U-WER). U-WER is measured on words that are NOT in the biasing list, while B-WER is measured on words that are IN the biasing list. For insertion errors, if the inserted phrase is in the biasing list, it is counted towards B-WER, otherwise it is counted towards U-WER.

\vspace{-0.15cm}
\subsection{Contextual ASR Accuracy}
\vspace{-0.05cm}

\begin{table}[!t]
\centering
\setlength{\belowcaptionskip}{0.05cm} 
\setlength{\abovecaptionskip}{0.15cm} 
\caption{Results of contextual phrase filtering on contextual model with biasing list size N = 1000}
\label{filter}
\resizebox{0.48\textwidth}{!}{
\begin{tabular}{l|cc|cc}
\hline
\multirow{2}{*}{\quad Model} & \multicolumn{2}{c|}{without filter}                                                                                     & \multicolumn{2}{c}{with filter}                                                                                        \\ \cline{2-5} 
                       & test-clean                                                & test-other                                                  & test-clean                                                 & test-other                                                 \\ \hline\hline
CTC + bias             & \begin{tabular}[c]{@{}c@{}}6.89\\ (6.0 / 14.7)\end{tabular} & \begin{tabular}[c]{@{}c@{}}14.85\\ (12.9 / 32.6)\end{tabular} & \begin{tabular}[c]{@{}c@{}}4.95\\ (3.72 / 15.3)\end{tabular} & \begin{tabular}[c]{@{}c@{}}10.81\\ (8.3 / 32.7)\end{tabular} \\ \hline
AED + bias             & \begin{tabular}[c]{@{}c@{}}3.81\\ (2.9 / 11.4)\end{tabular} & \begin{tabular}[c]{@{}c@{}}8.75\\ (6.9 / 25.3)\end{tabular}   & \begin{tabular}[c]{@{}c@{}}3.61\\ (2.6 / 12.1)\end{tabular}  & \begin{tabular}[c]{@{}c@{}}7.90\\ (5.9 / 25.8)\end{tabular}  \\ \hline
C-T + bias             & \begin{tabular}[c]{@{}c@{}}3.88\\ (2.9 / 11.9)\end{tabular} & \begin{tabular}[c]{@{}c@{}}8.28\\ (6.4 / 24.5)\end{tabular}   & \begin{tabular}[c]{@{}c@{}}3.76\\ (2.7 / 12.8)\end{tabular}   & \begin{tabular}[c]{@{}c@{}}7.96\\ (5.8 / 26.8)\end{tabular}   \\ \hline
\end{tabular}
}
\vspace{-0.2cm} 
\end{table}

\begin{table}[!t]
\centering
\setlength{\belowcaptionskip}{0.05cm} 
\setlength{\abovecaptionskip}{0.15cm} 
\caption{Comparison of proposed method and shallow fusion}
\label{fusion}
\resizebox{0.48\textwidth}{!}{
\setlength{\tabcolsep}{7mm}{
\begin{tabular}{l|c|c}
\hline
\quad \ \ Model           & test-clean                                                & test-other                                                \\ \hline\hline
AED baseline    & \begin{tabular}[c]{@{}c@{}}4.29\\ (2.6 / 18.3)\end{tabular} & \begin{tabular}[c]{@{}c@{}}9.16\\ (5.9 / 37.5)\end{tabular} \\ \hline
\qquad + SF        & \begin{tabular}[c]{@{}c@{}}4.19\\ (2.7 / 16.8)\end{tabular} & \begin{tabular}[c]{@{}c@{}}8.79\\ (6.0 / 33.1)\end{tabular} \\ \hline
\qquad + bias      & \begin{tabular}[c]{@{}c@{}}3.88\\ (2.8 / 13.0)\end{tabular} & \begin{tabular}[c]{@{}c@{}}8.47\\ (6.4 / 26.6)\end{tabular} \\ \hline
 \qquad ++ SF & \begin{tabular}[c]{@{}c@{}}3.86\\ (2.9 / 12.4)\end{tabular}  & \begin{tabular}[c]{@{}c@{}}8.45\\ (6.5 / 25.5)\end{tabular}  \\ \hline
\end{tabular}
}
}
\vspace{-0.5cm} 
\end{table}

In this section, we tested the contextual recognition models and their non-contextual baseline on all these types of models. As shown in Table \ref{librispeech}, the results of experiments containing a non-empty biasing list demonstrate a decrease in B-WER. The best results were achieved when using a biasing list of 100, yielding an average 40.5\% relative improvement of B-WER compared to the baseline. When the size of the biasing list increased to 1000, B-WER still showed a noticeable decrease, but the reduction was less significant, and U-WER was also impacted to some degree. This suggests that the model is sensitive to the size of the biasing list. We found that using more distractors during the training process, or not sampling contextual phrases for some utterances can reduce U-WER when using a large biasing list, but will also decrease the improvement of B-WER.

\vspace{-0.15cm}
\subsection{Contextual Phrase Filtering Performance}
\vspace{-0.05cm}

As shown in Table \ref{filter}, we tested the effect of contextual phrase filtering on all these types of models with a biasing list of size 1000. After applying the filtering method, the impact of the biasing list on U-WER was significantly reduced and largely regressed to the level when an empty biasing list was used, while B-WER showed only a slight increase compared to the results without contextual phrase filtering.

\vspace{-0.15cm}
\subsection{Comparison with Shallow Fusion}
\vspace{-0.05cm}

In order to compare the proposed deep bias method and shallow fusion, we separately tested the effect of using each method alone and using them together on the contextualized AED model. To ensure fairness, we extracted the contextual phrases in each utterance in the test set into a list, with a list size of 4250 for the \textit{test-clean} set and 3838 for the \textit{test-other} set. Two methods use the same biasing list. When evaluating the results of the shallow fusion, we used the shallow fusion method based on OTF rescoring described in~\cite{wenet}. As shown in Table \ref{fusion}, the shallow fusion method reduced the average B-WER by 6.9\% relative to the baseline and the proposed deep bias method reduced the average B-WER by 29.2\%. Compared to shallow fusion, the improvement in B-WER from deep bias was apparently greater, and the use of both methods together can further reduce the B-WER, although the impact on U-WER will also accumulate.

\vspace{-0.15cm}
\subsection{Ablation Study}
\vspace{-0.05cm}

\begin{table}[!t]
\centering
\setlength{\belowcaptionskip}{0.05cm} 
\setlength{\abovecaptionskip}{0.15cm} 
\caption{Results of ablation study}
\label{ablation}
\resizebox{0.48\textwidth}{!}{
\begin{tabular}{l|cc|cc}
\hline
\multirow{2}{*}{\quad Model}                                           & \multicolumn{2}{c|}{N=100}                                                                                            & \multicolumn{2}{c}{N=1000}                                                                                           \\ \cline{2-5} 
                                                                 & test-clean                                                & test-other                                                & test-clean                                                & test-other                                                \\ \hline\hline
AED + bias                                                       & \begin{tabular}[c]{@{}c@{}}\textbf{3.40}\\ \textbf{(2.6 / 10.4)}\end{tabular} & \begin{tabular}[c]{@{}c@{}}7.77\\ (6.0 / 23.0)\end{tabular} & \begin{tabular}[c]{@{}c@{}}\textbf{3.81}\\ \textbf{(2.9 / 11.4)}\end{tabular} & \begin{tabular}[c]{@{}c@{}}\textbf{8.75}\\ \textbf{(6.9 / 25.3)}\end{tabular} \\ \hline
\begin{tabular}[c]{@{}c@{}} \qquad - share\end{tabular} & \begin{tabular}[c]{@{}c@{}}3.50\\ (2.7 / 10.2)\end{tabular} & \begin{tabular}[c]{@{}c@{}}\textbf{7.68}\\ \textbf{(6.2 / 21.0)}\end{tabular} & \begin{tabular}[c]{@{}c@{}}4.13\\ (3.3 / 11.3)\end{tabular} & \begin{tabular}[c]{@{}c@{}}9.43\\ (7.7 / 24.4)\end{tabular} \\ \hline
\begin{tabular}[c]{@{}c@{}} \qquad - loss\end{tabular}    & \begin{tabular}[c]{@{}c@{}}4.26\\ (2.6 / 18.0)\end{tabular} & \begin{tabular}[c]{@{}c@{}}9.31\\ (5.9 / 39.0)\end{tabular} & \begin{tabular}[c]{@{}c@{}}4.26\\ (2.6 / 18.0)\end{tabular} & \begin{tabular}[c]{@{}c@{}}9.31\\ (5.9 / 39.0)\end{tabular} \\ \hline\hline
C-T + bias                                                       & \begin{tabular}[c]{@{}c@{}}\textbf{3.66}\\ \textbf{(2.8 / 11.2)}\end{tabular} & \begin{tabular}[c]{@{}c@{}}\textbf{7.63}\\ \textbf{(6.0 / 22.1)}\end{tabular} & \begin{tabular}[c]{@{}c@{}}\textbf{3.88}\\ \textbf{(2.9 / 11.9)}\end{tabular} & \begin{tabular}[c]{@{}c@{}}\textbf{8.28}\\ \textbf{(6.4 / 24.5)}\end{tabular} \\ \hline
\begin{tabular}[c]{@{}c@{}} \qquad - loss\end{tabular}    & \begin{tabular}[c]{@{}c@{}}4.05\\ (2.8 / 14.5)\end{tabular} & \begin{tabular}[c]{@{}c@{}}8.65\\ (6.2 / 30.5)\end{tabular} & \begin{tabular}[c]{@{}c@{}}4.24\\ (2.9 / 15.4)\end{tabular} & \begin{tabular}[c]{@{}c@{}}8.99\\ (6.3 / 32.8)\end{tabular} \\ \hline
\end{tabular}
}
\vspace{-0.2cm} 
\end{table}

\begin{table}[!t]
\centering
\setlength{\belowcaptionskip}{0.05cm} 
\setlength{\abovecaptionskip}{0.15cm} 
\caption{Results of contextualized AED model on Earnings21}
\label{gigaspeech}
\resizebox{0.48\textwidth}{!}{
\setlength{\tabcolsep}{3mm}{
\renewcommand\arraystretch{1.6}
\scriptsize
\begin{tabular}{l|cccc}
\hline
\quad \ \ Model    & WER   & Recall & Precision & F1    \\ \hline\hline
AED Baseline & 14.96 & 58.66  & 91.52     & 71.50 \\ \hline
AED + bias   & 15.12 & 79.11  & 80.67     & 79.88 \\ \hline
\end{tabular}
}
}
\vspace{-0.5cm} 
\end{table}

Table \ref{ablation} presents the results of ablation study. We tested the performance of the contextualized AED model without sharing the CTC linear parameters in the CPP Network and without the addition of the bias loss. The results of not sharing the CTC linear parameters showed better B-WER performance, but the U-WER was more sensitive to the size of the biasing list. When a larger biasing list was used, there was a noticeable increase in the WER loss. Without the bias loss, the contextualized AED model completely lost its ability to perform bias recognition. In the contextualized Transducer, when the bias loss is removed, the resulting model is essentially CATT, which had the ability to perform bias recognition, but the improvement in the B-WER was not as significant as before.

\vspace{-0.15cm}
\subsection{Test on Earnings21}
\vspace{-0.05cm}

We trained a contextualized AED model on the GigaSpeech dataset~\cite{gigaspeech} and evaluated it on the Earnings21 dataset~\cite{earnings}, with results presented in Table \ref{gigaspeech}. Similar to~\cite{earnings}, we demonstrate the WER and phrase-level recall, precision, and F1 scores of two models on the Earnings21 dataset. When using the contextualized AED model and applying contextual phrase filtering, the recall of contextual phrases is significantly improved, resulting in a relative improvement of 33.2\% compared to the baseline. Although the precision was impacted to a certain extent, the overall F1 score still improves by 11.7\% relative. Due to the small proportion of biased phrases in the test set, contextual biasing has a minimal impact on overall WER.

\vspace{-0.2cm}
\section{Conclusions}

In this paper, we propose a deep biasing method based on the contextual phrase prediction network. The context phrase prediction network predicts context phrases contained in the utterances and calculates CTC loss on the prediction results as an auxiliary loss for the training of the contextualized model. Compared with previous works, our approach achieves significant B-WER improvement on all CTC, AED, and Transducer models. In addition, we apply a context phrase filtering method to the final context-aware ASR model, which effectively reduces the U-WER loss when using a larger biasing list.


\nocite{*}
\bibliography{mybib.bib} 

\begin{thebibliography}{10}
\providecommand{\url}[1]{#1}
\csname url@samestyle\endcsname
\providecommand{\newblock}{\relax}
\providecommand{\bibinfo}[2]{#2}
\providecommand{\BIBentrySTDinterwordspacing}{\spaceskip=0pt\relax}
\providecommand{\BIBentryALTinterwordstretchfactor}{4}
\providecommand{\BIBentryALTinterwordspacing}{\spaceskip=\fontdimen2\font plus
\BIBentryALTinterwordstretchfactor\fontdimen3\font minus
  \fontdimen4\font\relax}
\providecommand{\BIBforeignlanguage}[2]{{%
\expandafter\ifx\csname l@#1\endcsname\relax
\typeout{** WARNING: IEEEtran.bst: No hyphenation pattern has been}%
\typeout{** loaded for the language `#1'. Using the pattern for}%
\typeout{** the default language instead.}%
\else
\language=\csname l@#1\endcsname
\fi
#2}}
\providecommand{\BIBdecl}{\relax}
\BIBdecl

\bibitem{2006ctc}
A.~Graves, S.~Fern{\'{a}}ndez, F.~J. Gomez, and J.~Schmidhuber, ``Connectionist
  temporal classification: labelling unsegmented sequence data with recurrent
  neural networks,'' in \emph{Machine Learning, Proceedings of the Twenty-Third
  International Conference {(ICML} 2006)}, W.~W. Cohen and A.~W. Moore,
  Eds.\hskip 1em plus 0.5em minus 0.4em\relax {ACM}, 2006, pp. 369--376.

\bibitem{2014ctc}
A.~Graves and N.~Jaitly, ``Towards end-to-end speech recognition with recurrent
  neural networks,'' in \emph{Proceedings of the 31th International Conference
  on Machine Learning, {ICML} 2014}.\hskip 1em plus 0.5em minus 0.4em\relax
  JMLR.org, 2014, pp. 1764--1772.

\bibitem{2019rnnt}
Y.~He, T.~N. Sainath, R.~Prabhavalkar, I.~McGraw, R.~Alvarez, D.~Zhao,
  D.~Rybach, A.~Kannan, Y.~Wu, R.~Pang, Q.~Liang, D.~Bhatia, Y.~Shangguan,
  B.~Li, G.~Pundak, K.~C. Sim, T.~Bagby, S.~Chang, K.~Rao, and A.~Gruenstein,
  ``Streaming end-to-end speech recognition for mobile devices,'' in
  \emph{{IEEE} International Conference on Acoustics, Speech and Signal
  Processing, {ICASSP} 2019}.\hskip 1em plus 0.5em minus 0.4em\relax {IEEE},
  2019, pp. 6381--6385.

\bibitem{2021rnnt}
Y.~Zhang, S.~Sun, and L.~Ma, ``Tiny transducer: {A} highly-efficient speech
  recognition model on edge devices,'' in \emph{{IEEE} International Conference
  on Acoustics, Speech and Signal Processing, {ICASSP} 2021}.\hskip 1em plus
  0.5em minus 0.4em\relax {IEEE}, 2021, pp. 6024--6028.

\bibitem{2015aed}
J.~Chorowski, D.~Bahdanau, D.~Serdyuk, K.~Cho, and Y.~Bengio, ``Attention-based
  models for speech recognition,'' in \emph{Advances in Neural Information
  Processing Systems 28: Annual Conference on Neural Information Processing
  Systems 2015}, C.~Cortes, N.~D. Lawrence, D.~D. Lee, M.~Sugiyama, and
  R.~Garnett, Eds., 2015, pp. 577--585.

\bibitem{2014aed}
K.~Cho, B.~van Merrienboer, {\c{C}}.~G{\"{u}}l{\c{c}}ehre, D.~Bahdanau,
  F.~Bougares, H.~Schwenk, and Y.~Bengio, ``Learning phrase representations
  using {RNN} encoder-decoder for statistical machine translation,'' in
  \emph{Proceedings of the 2014 Conference on Empirical Methods in Natural
  Language Processing, {EMNLP} 2014}, A.~Moschitti, B.~Pang, and W.~Daelemans,
  Eds.\hskip 1em plus 0.5em minus 0.4em\relax {ACL}, 2014, pp. 1724--1734.

\bibitem{2016aed}
W.~Chan, N.~Jaitly, Q.~V. Le, and O.~Vinyals, ``Listen, attend and spell: {A}
  neural network for large vocabulary conversational speech recognition,'' in
  \emph{2016 {IEEE} International Conference on Acoustics, Speech and Signal
  Processing, {ICASSP} 2016}.\hskip 1em plus 0.5em minus 0.4em\relax {IEEE},
  2016, pp. 4960--4964.

\bibitem{2018sf}
I.~Williams, A.~Kannan, P.~S. Aleksic, D.~Rybach, and T.~N. Sainath,
  ``Contextual speech recognition in end-to-end neural network systems using
  beam search,'' in \emph{Interspeech 2018, 19th Annual Conference of the
  International Speech Communication Association}, B.~Yegnanarayana, Ed.\hskip
  1em plus 0.5em minus 0.4em\relax {ISCA}, 2018, pp. 2227--2231.

\bibitem{2019sf}
Z.~Chen, M.~Jain, Y.~Wang, M.~L. Seltzer, and C.~Fuegen, ``End-to-end
  contextual speech recognition using class language models and a token passing
  decoder,'' in \emph{{IEEE} International Conference on Acoustics, Speech and
  Signal Processing, {ICASSP} 2019, Brighton, United Kingdom, May 12-17,
  2019}.\hskip 1em plus 0.5em minus 0.4em\relax {IEEE}, 2019, pp. 6186--6190.

\bibitem{2019sf2}
D.~Zhao, T.~N. Sainath, D.~Rybach, P.~Rondon, D.~Bhatia, B.~Li, and R.~Pang,
  ``Shallow-fusion end-to-end contextual biasing,'' in \emph{Interspeech 2019,
  20th Annual Conference of the International Speech Communication
  Association}, G.~Kubin and Z.~Kacic, Eds.\hskip 1em plus 0.5em minus
  0.4em\relax {ISCA}, 2019, pp. 1418--1422.

\bibitem{2020sf}
R.~Huang, O.~Abdel{-}Hamid, X.~Li, and G.~Evermann, ``Class {LM} and word
  mapping for contextual biasing in end-to-end {ASR},'' in \emph{Interspeech
  2020, 21st Annual Conference of the International Speech Communication
  Association}, H.~Meng, B.~Xu, and T.~F. Zheng, Eds.\hskip 1em plus 0.5em
  minus 0.4em\relax {ISCA}, 2020, pp. 4348--4351.

\bibitem{2021sf}
S.~Kim, Y.~Shangguan, J.~Mahadeokar, A.~Bruguier, C.~Fuegen, M.~L. Seltzer, and
  D.~Le, ``Improved neural language model fusion for streaming recurrent neural
  network transducer,'' in \emph{{IEEE} International Conference on Acoustics,
  Speech and Signal Processing, {ICASSP} 2021}.\hskip 1em plus 0.5em minus
  0.4em\relax {IEEE}, 2021, pp. 7333--7337.

\bibitem{2019nnbias}
Z.~Chen, M.~Jain, Y.~Wang, M.~L. Seltzer, and C.~Fuegen, ``Joint grapheme and
  phoneme embeddings for contextual end-to-end {ASR},'' in \emph{Interspeech
  2019, 20th Annual Conference of the International Speech Communication
  Association}, G.~Kubin and Z.~Kacic, Eds.\hskip 1em plus 0.5em minus
  0.4em\relax {ISCA}, 2019, pp. 3490--3494.

\bibitem{2021nnbias}
M.~Han, L.~Dong, S.~Zhou, and B.~Xu, ``Cif-based collaborative decoding for
  end-to-end contextual speech recognition,'' in \emph{{IEEE} International
  Conference on Acoustics, Speech and Signal Processing, {ICASSP} 2021}.\hskip
  1em plus 0.5em minus 0.4em\relax {IEEE}, 2021, pp. 6528--6532.

\bibitem{2021nnbias2}
G.~Sun, C.~Zhang, and P.~C. Woodland, ``Tree-constrained pointer generator for
  end-to-end contextual speech recognition,'' in \emph{{IEEE} Automatic Speech
  Recognition and Understanding Workshop, {ASRU} 2021}.\hskip 1em plus 0.5em
  minus 0.4em\relax {IEEE}, 2021, pp. 780--787.

\bibitem{2021nnbias3}
D.~Le, M.~Jain, G.~Keren, S.~Kim, Y.~Shi, J.~Mahadeokar, J.~Chan, Y.~Shangguan,
  C.~Fuegen, O.~Kalinli, Y.~Saraf, and M.~L. Seltzer, ``Contextualized
  streaming end-to-end speech recognition with trie-based deep biasing and
  shallow fusion,'' in \emph{Interspeech 2021, 22nd Annual Conference of the
  International Speech Communication Association}, H.~Hermansky,
  H.~Cernock{\'{y}}, L.~Burget, L.~Lamel, O.~Scharenborg, and
  P.~Motl{\'{\i}}cek, Eds.\hskip 1em plus 0.5em minus 0.4em\relax {ISCA}, 2021,
  pp. 1772--1776.

\bibitem{2022nnbias}
S.~Dingliwal, M.~Sunkara, S.~Ronanki, J.~Farris, K.~Kirchhoff, and S.~Bodapati,
  ``Personalization of {CTC} speech recognition models,'' in \emph{{IEEE}
  Spoken Language Technology Workshop, {SLT} 2022}.\hskip 1em plus 0.5em minus
  0.4em\relax {IEEE}, 2022, pp. 302--309.

\bibitem{2018nnbias}
G.~Pundak, T.~N. Sainath, R.~Prabhavalkar, A.~Kannan, and D.~Zhao, ``Deep
  context: End-to-end contextual speech recognition,'' in \emph{2018 {IEEE}
  Spoken Language Technology Workshop, {SLT} 2018}.\hskip 1em plus 0.5em minus
  0.4em\relax {IEEE}, 2018, pp. 418--425.

\bibitem{2021nnbias4}
F.~Chang, J.~Liu, M.~Radfar, A.~Mouchtaris, M.~Omologo, A.~Rastrow, and
  S.~Kunzmann, ``Context-aware transformer transducer for speech recognition,''
  in \emph{{IEEE} Automatic Speech Recognition and Understanding Workshop,
  {ASRU} 2021}.\hskip 1em plus 0.5em minus 0.4em\relax {IEEE}, 2021, pp.
  503--510.

\bibitem{filter}
Z.~Yang, S.~Sun, X.~Wang, Y.~Zhang, L.~Ma, and L.~Xie, ``Two stage contextual
  word filtering for context bias in unified streaming and non-streaming
  transducer,'' \emph{CoRR}, vol. abs/2301.06735, 2023.

\bibitem{librispeech}
V.~Panayotov, G.~Chen, D.~Povey, and S.~Khudanpur, ``Librispeech: An {ASR}
  corpus based on public domain audio books,'' in \emph{2015 {IEEE}
  International Conference on Acoustics, Speech and Signal Processing, {ICASSP}
  2015}.\hskip 1em plus 0.5em minus 0.4em\relax {IEEE}, 2015, pp. 5206--5210.

\bibitem{specaugment}
D.~S. Park, W.~Chan, Y.~Zhang, C.~Chiu, B.~Zoph, E.~D. Cubuk, and Q.~V. Le,
  ``Specaugment: {A} simple data augmentation method for automatic speech
  recognition,'' in \emph{Interspeech 2019, 20th Annual Conference of the
  International Speech Communication Association}, G.~Kubin and Z.~Kacic,
  Eds.\hskip 1em plus 0.5em minus 0.4em\relax {ISCA}, 2019, pp. 2613--2617.

\bibitem{gigaspeech}
G.~Chen, S.~Chai, G.~Wang, J.~Du, W.~Zhang, C.~Weng, D.~Su, D.~Povey, J.~Trmal,
  J.~Zhang, M.~Jin, S.~Khudanpur, S.~Watanabe, S.~Zhao, W.~Zou, X.~Li, X.~Yao,
  Y.~Wang, Z.~You, and Z.~Yan, ``Gigaspeech: An evolving, multi-domain {ASR}
  corpus with 10, 000 hours of transcribed audio,'' in \emph{Interspeech 2021,
  22nd Annual Conference of the International Speech Communication
  Association}, H.~Hermansky, H.~Cernock{\'{y}}, L.~Burget, L.~Lamel,
  O.~Scharenborg, and P.~Motl{\'{\i}}cek, Eds.\hskip 1em plus 0.5em minus
  0.4em\relax {ISCA}, 2021, pp. 3670--3674.

\bibitem{earnings}
J.~D. Fox and N.~Delworth, ``Improving contextual recognition of rare words
  with an alternate spelling prediction model,'' in \emph{Interspeech 2022,
  23rd Annual Conference of the International Speech Communication
  Association}, H.~Ko and J.~H.~L. Hansen, Eds.\hskip 1em plus 0.5em minus
  0.4em\relax {ISCA}, 2022, pp. 3914--3918.

\bibitem{wenet}
B.~Zhang, D.~Wu, Z.~Peng, X.~Song, Z.~Yao, H.~Lv, L.~Xie, C.~Yang, F.~Pan, and
  J.~Niu, ``Wenet 2.0: More productive end-to-end speech recognition toolkit,''
  in \emph{Interspeech 2022, 23rd Annual Conference of the International Speech
  Communication Association}, H.~Ko and J.~H.~L. Hansen, Eds.\hskip 1em plus
  0.5em minus 0.4em\relax {ISCA}, 2022, pp. 1661--1665.

\end{thebibliography}
\bibliographystyle{IEEEtran} 

\end{document}